\documentclass{emulateapj}
\usepackage{apjfonts}

\newcommand{\dec}{\rm{dec}}
\newcommand{\iso}{\rm{iso}}
\newcommand{\br}{\rm{br}}
\newcommand{\ex}{\rm{ex}}
\slugcomment{ApJ accepted 2006 December 30}
\shorttitle{A collapsar jet model of GRB 060218}
\shortauthors{Toma et al.}
\begin{document}
\title{
Low-Luminosity GRB 060218: A Collapsar Jet from a Neutron Star,
Leaving a Magnetar as a Remnant?
}
\author{Kenji Toma\altaffilmark{1}, Kunihito Ioka\altaffilmark{1}, 
Takanori Sakamoto\altaffilmark{2,3}, and Takashi Nakamura\altaffilmark{1}}

\altaffiltext{1}{Department of Physics, Kyoto University,
Kyoto 606-8502, Japan;
toma@tap.scphys.kyoto-u.ac.jp}
\altaffiltext{2}{NASA - Goddard Space Flight Center, 
Greenbelt, Maryland 20771, USA;
takanori@milkyway.gsfc.nasa.gov}
\altaffiltext{3}{National Research Council, 2101 Constitution
Ave NW, Washington DC 20418, USA}
\begin{abstract}

The gamma-ray burst (GRB) 060218 has $\sim 10^5$ times lower luminosity 
than typical long GRBs, and is associated with a supernova (SN).
The radio afterglow displays no jet break,
so that this burst might arise from a mildly-relativistic 
spherical outflow produced by the SN shock sweeping the stellar surface. 
Since this model is energetically difficult,
we propose that the radio afterglow is produced by
a non-relativistic phase of an initially collimated outflow (jet).
Our jet model is supported by the detection of optical linear polarization 
in the SN component.
We also show analytically that the jet can penetrate a progenitor star.
We analyzed the observational data of the prompt emission of this burst and
obtained a smooth power-law light curve which might last longer than $10^6$~s.
This behavior contrasts with the long intermittent activities with the 
X-ray flares of typical GRBs, implying that the central engine 
of this burst is different from those of typical GRBs.
This argument is consistent with the analysis of the SN component of this 
burst, which suggests that the progenitor star was less massive and collapsed
to a neutron star instead of a black hole.
The collimation-corrected event rate of such low-luminosity GRBs is 
estimated to be $\sim 10$ times higher than that of typical long GRBs, 
and they might form a different GRB population:
low-luminosity GRBs are produced by mildly-relativistic 
jets from neutron stars at the collapses of massive stars, 
while typical long GRBs by highly-relativistic jets from 
black holes.
We suggest that the central engine of GRB 060218 is
a pulsar (or a magnetar) with the initial rotation period $P_0 \sim 10$~ms
and the magnetic field $B \sim 10^{16}$~G. 
A giant flare from the magnetar might be observed in future. 

\end{abstract}


\section{Introduction}
\label{sec:intro}

The gamma-ray burst (GRB) 060218 is the second nearest event
($z=0.033$) and it is spectroscopically associated with the supernova (SN) 2006aj
\citep{modjaz06,sollerman06,pian06,mazzali06,mirabal06,cobb06,ferrero06}.
Such a long GRB/SN association is common
and this event further supports the established picture
that all long GRBs are related to the deaths
of massive stars \citep[for recent reviews,][]{woosley06,meszaros06,piran05,zhang04}.
In the most popular model of long GRBs, the so-called collapsar model,
the core of a massive star
collapses to a black hole or a neutron star, which drives a highly relativistic jet,
breaking out the star and making a GRB \citep{woosley93,macfadyen99}.
The relativistic speed of the outflow is required for the non-thermal prompt
emission \citep[][and references therein]{lithwick01}.
The collimation of the outflow is strongly suggested by 
a break in the afterglow, since it is produced by the sideways expansion
of the jet \citep[e.g.,][]{rhoads99,sari99,harrison99}.

However, the prompt and afterglow emission of GRB 060218 have many peculiarities
\citep{campana06,soderberg06b,ghisellini06b,butler06,liang06c}:
\begin{itemize}
\item[(1)] The duration of the prompt non-thermal emission in the high energy 
band ($15-150$~keV) is $\delta t \sim 10^3$~s, and 
thus this event is one of the longest bursts.
\item[(2)] The isotropic-equivalent luminosity of the prompt emission is extremely 
low $\sim 10^{47}~\rm{erg}~\rm{s}^{-1}$, which is about $10^5$ times
lower than those of typical cosmological GRBs.
The isotropic-equivalent energy, extrapolated to the $1-10^4$~keV band
in the central engine frame, is 
$E_{\gamma,\iso} \simeq 6\times10^{49}$~erg.
\item[(3)] The spectrum of the prompt emission is quite soft compared to those of 
typical bright GRBs, and the averaged spectral peak energy in the central engine frame 
is $\simeq 4.9$~keV.
(Note that this event obeys the Amati correlation, despite the other peculiarities.
See \S~\ref{sec:prompt} for more details.)
\item[(4)] Thermal components are detected in the X-ray band at $t \lesssim 10^4$~s
and in the UV/optical band at $10^4 \lesssim t \lesssim 10^5$~s,
while other GRBs do not exhibit such a clear thermal component.
\item[(5)] The X-ray, UV/optical, and radio afterglows show chromatic features.
The X-ray afterglow (at $t \gtrsim 10^4$~s) decays with a standard temporal slope, but
has a spectrum much steeper than those of typical GRB X-ray afterglows.
The UV/optical afterglow is quite dim and dominated by the thermal component and
the SN component.
Only the radio afterglow seems rather typical and explainable in the standard 
external shock synchrotron model \citep{sari98,rees92,paczynski93}, 
but does not show a jet break until $t \simeq 22$~days.
\end{itemize}

Although such a low-luminosity event looks rare, 
the intrinsic event rate 
could be very high $R_{\rm LL} \sim 10^2~{\rm Gpc}^{-3}~{\rm yr}^{-1}$
compared with the local rate of typical long GRBs deduced from the BATSE data, 
$R_{\rm LG} \sim 1~{\rm Gpc}^{-3}~{\rm yr}^{-1}$
\citep{soderberg06b,pian06,cobb06,liang06b,guetta05}.
For this reason, it has been actively debated 
whether low-luminosity GRBs form a new GRB population
and whether they have intrinsically different outflow mechanisms and emission 
mechanisms
\citep[see][]{stanek06,ghisellini06a,kaneko06,amati06,stratta06,dai06,wang06}.
It has been also argued that the high energy neutrino background from the 
low-luminosity GRBs could be comparable with or larger than that from typical
long GRBs \citep{murase06,gupta06}.

For GRB 060218, it has been widely suggested that the outflow is spherical
since the radio afterglow has no jet break \citep{soderberg06b,fan06}.
If true, the collapsar model cannot be applied to this event because
the outflow becomes non-relativistic by loading all the matter of a progenitor star.
The relativistic spherical outflow might be produced 
by the outermost parts of the stellar envelope that the SN shock accelerates when 
propagating through the steep density gradient near the stellar surface 
\citep{colgate74,matzner99,tan01}.
However, \citet{tan01} has shown that 
the energy $\sim 10^{48}$~erg is transferred to mildly relativistic material
when the kinetic energy of the SN is $E_{\rm SN} \sim 10^{52}$~erg.
For GRB 060218, $E_{\rm SN}$ is estimated as $\simeq 2\times10^{51}$~erg
\citep{mazzali06}, so that it is quite unlikely
that the prompt non-thermal emission with 
$E_{\gamma,\iso} \simeq 6\times10^{49}$~erg
is produced by this type of the outflow \citep[see also][]{matzner03}.
\citet{li06} has also shown that 
the energy of the thermal components $\gtrsim 10^{49}$~erg
is too large to be explained by the shock breakout in the underlying SN.

In this paper, we show that GRB 060218 can be produced
by the standard collapsar jet model.
We show that the available radio data may be interpreted 
as a non-relativistic phase 
of the external shock {\it after the jet break} within the standard model 
\citep[e.g.,][]{frail00,livio00}.
We argue that the outflow with an initial opening angle $\theta_0 \simeq 0.3$
and Lorentz factor $\Gamma_0 \simeq 5$ can produce the synchrotron radiation which 
explains the radio afterglow and is compatible with the UV/optical and
X-ray afterglow (\S~\ref{sec:afterglow}).
We also examine whether such a wide and weak jet can penetrate a progenitor star
by extending analytical considerations of the collapsar model by
\citet{matzner03} (\S~\ref{sec:collapsar}).

Remarkably, the detection of optical linear polarization in the SN component 
of this event has been recently reported \citep{gorosabel06}.
This observation strongly supports our
arguments that GRB 060218 arises from a jet.

Within the jet scenario, there is a possibility that low-luminosity GRBs arise as 
typical cosmological GRB jets viewed off-axis 
\citep[e.g.,][]{yama03,ramirez05,granot05,toma05}. 
The relativistic jet emits $\gamma$-rays into the expansion direction
through the beaming effect.
Thus, if the jet is viewed off-axis,
the $\gamma$-ray flux is strongly suppressed.
This scenario leads to similar event rates of typical GRBs and
low-luminosity GRBs, which seems inconsistent with the observations
\citep{cobb06}.
Aside from this statistical argument, we also examine the off-axis scenario for
GRB 060218 and conclude that this scenario is unlikely for this event 
because an unrealistically high $\gamma$-ray efficiency is required.

In addition,
we analyzed the data of 
the Burst Alert Telescope \citep[BAT;][]{barthelmy05}
and the X-Ray Telescope \citep[XRT;][]{burrows05a} 
of {\it Swift} satellite \citep{gehrels04}.
We found that the non-thermal component of the prompt emission of GRB 060218 
may be fitted by the Band function with the spectral parameters similar to those of 
typical GRBs.
Furthermore, we found that the light curve of the non-thermal component evolves 
smoothly as a power-law function of time, and it might last so long duration as to
connect to the anomalous X-ray afterglow detected up to $\sim 10^6$~s.
This behavior contrasts with the long intermittent activities with  
the X-ray flares of typical GRBs, which might be a sign of the difference
of the central engines.
This is also suggested by the analysis of the SN component of this event.
The SN spectrum has less broad lines than
those of other GRB-SNe and lacks oxygen lines, and the light curve evolves
somewhat faster.
All these facts indicate that the total kinetic energy $E_{\rm SN}$
and the ejected mass $M_{\rm ej}$ are both less than those of other GRB-SNe.
\citet{mazzali06} performed a detailed modeling of the spectra and light
curve of the SN component and obtained $E_{\rm SN} \simeq 2\times10^{51}$~erg 
and $M_{\rm ej} \simeq 2 M_{\odot}$.
Then they have suggested that the progenitor star of this burst was less massive 
than those of typical long GRBs and collapsed to a neutron star 
instead of a black hole.

Our goal is showing that the low-luminosity GRB 060218 has a prompt non-thermal 
emission with a typical Band spectrum and may originate from a standard
collapsar jet possibly driven by a neutron star.
We will not discuss the emission mechanism of the prompt non-thermal emission 
and the thermal emissions. 
This paper is organized as follows.
In \S~\ref{sec:prompt}, we show the results of our analysis of the BAT and XRT data
and suggest several implications for the nature of the prompt non-thermal emission.
In \S~\ref{sec:afterglow}, we display a jet model of this event, and
discuss whether the jet can make a hole in the star
in \S~\ref{sec:collapsar}.
A summary and discussion are given in \S~\ref{sec:discussion}.
Since GRB 060218 is very close ($z=0.033$),
we will neglect the cosmological effect, i.e., we set $z=0$, for
simplicity throughout this paper.

\section{The Prompt Non-thermal Emission}
\label{sec:prompt}

\subsection{BAT and XRT data analysis}

We performed the analysis of the BAT and XRT data using the standard 
Swift software package (HEAsoft 6.0.4) and the CALDB 2005-11-28.  The 
detector plane histogram (DPH) data were used to extract the spectra of 
BAT.  Before producing the spectra, we applied {\tt baterebin} to 
rebin the DPH data using the most accurate non-linear energy correction.  
The detector map for 
disabling noisy detectors in the analysis was created by {\tt bathotpix}.  
The mask-weighting map was created by {\tt batmaskwtimg} using the
optical afterglow position.  
With including the detector map and the mask-weighting map to {\tt
batbinevt}, 13 BAT spectra were extracted from the DPH file which 
contains the data just after the spacecraft slew.  The BAT response
matrices were generated by {\tt batdrmgen}.  The systematic error vectors 
were applied to the BAT spectra using {\tt batphasyserr}.  For the XRT data, we
obtained the cleaned event file in the window timing (WT) mode from 
the Swift HEASARC Archive.  The foreground was excluded by the box region of 
1.2$^{\prime} \times 0.6^{\prime}$.  The background region was 
extracted by the same region size, but outside of the foreground source 
region.  13 XRT foreground and background spectra using the same 
time intervals of the BAT spectra were extracted.  The auxiliary 
response file was generated by {\tt xrtmkarf}.  The XRT spectra were 
binned to contain a minimum of 20 photons for each spectral bin.  

The XRT and BAT spectral data were analyzed jointly with XSPEC 11.3.2.  
The energy ranges used in the analysis were 0.5--10 keV and 14.0--150
keV for XRT and BAT, respectively.  
We multiplied the constant factor to the spectral 
model to take into account the calibration uncertainty in the response 
matrices of the instruments.  As reported by the several authors 
\citep{kaneko06,campana06,butler06,liang06c}, 
the blackbody component was necessary to obtain 
a good fit for the low energy part of the XRT spectrum.  Thus, 
we performed the fitting with an additional blackbody (BB) component to
an absorbed Band function and an absorbed power-law 
times exponential cutoff (CPL) model.
For the purpose of this paper, we are interested in the spectral parameters
of the non-thermal component.
To obtain the well constrained spectral parameters of the non-thermal component, 
we fixed the parameters of the absorption ($N_{\rm H}$) and the BB component 
to their best fit values of each fit, and then calculate the uncertainties of 
the spectral parameters of the non-thermal component.  
The results are shown in Table~\ref{tab:band} and \ref{tab:cpl}.
The most of the spectra were well fitted with an
absorbed CPL model with an additional BB component as reported by 
\citet{kaneko06} and \citet{campana06}.  
However, we found a significant improvement in a fit with the Band 
function for the last four time intervals from 1550 s 
to 2732 s after the BAT trigger.  The differences in $\chi^{2}$ 
between the CPL and Band fit were 9.6, 11.7, 17.0, and 13.8 in 1 degree of freedom 
for these spectra.  Based on this result, we decided to adopt the 
spectral parameters derived by the Band function for all time intervals.  
The temperature of the BB component varies slightly between 0.12 and 0.29~keV.
This behavior is consistent with the result of the other analyses
\citep{kaneko06,campana06,butler06}.

\begin{figure}
\plotone{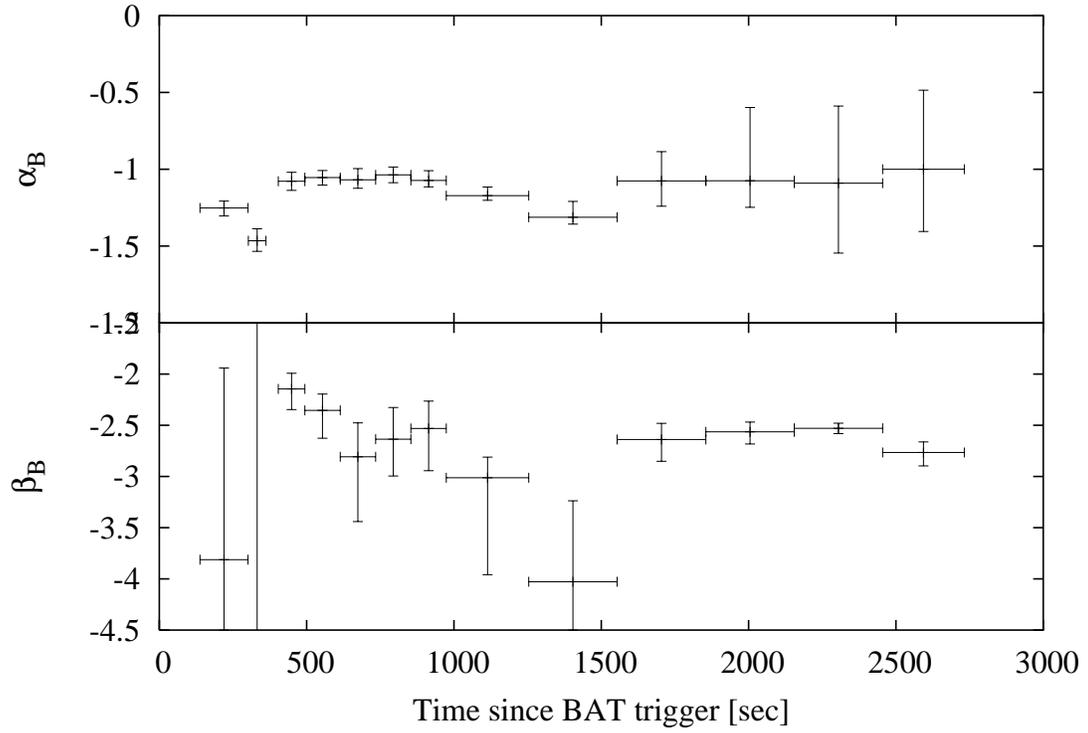}
\caption{
The low-energy photon index $\alpha_B$ and the high-energy photon index  
$\beta_B$ of the prompt non-thermal emission of GRB 060218 
as a function of time,
from the joint fit of the BAT and XRT data by the absorbed Band function plus
the blackbody function.
}
\label{fig:ab}
\end{figure}

Figure~\ref{fig:ab} shows the result of the temporal variations of 
the low-energy and high-energy photon index, $\alpha_B$ and $\beta_B$.
Figure~\ref{fig:fluxep_l} shows the light curve in the $15-150$~keV band
and the spectral peak energy $E_p$ as a function of time.
At the flux decay phase ($t \gtrsim 500$~s), the light curve and spectral 
peak energy are well described by power-law functions of time, 
$F \propto t^{-2.0}$ and $E_p \propto t^{-1.6}$ ({\it dotted lines}).

Investigating the consistency of the BAT light curves presented by us and 
\citet{campana06}, we found systematic differences especially for $t > 1000$~s.
Although both of the BAT light curves are based on the joint spectral analysis of 
the BAT and XRT data, \citet{campana06} use the CPL model while we use the Band
function.
We noticed that the fluxes of the light curve of \citet{campana06} at $t > 1000$~s
are in the level of $\sim 10$~mCrab.  
BAT can detect the Crab nebula at $6 \sigma$ level in 1~s exposure.  
If the BAT sensitivity can be scaled as a square root of the exposure time 
\citep{markwardt05}, we need the exposure time of $\sim 10^4$~s to detect the 
$10$~mCrab source assuming the Crab-like spectrum. 
On the other hand, the fluxes of our light curve are in the level of 
$\sim 100$~mCrab at the late phase.  
This flux level is reasonable to detect in $\sim 100$~s exposure by BAT. 
Thus, we believe that there is a systematic error by using the CPL model in the 
BAT flux calculation presented by \citet{campana06}. 
We also believe that this investigation strengthens our conclusion that 
the Band function is the best represented spectral model of the non-thermal emission of 
GRB 060218.  

\subsection{Implications for the nature of the prompt non-thermal emission} 

\subsubsection{Overall spectral properties}

We find that the prompt non-thermal emission of this event has spectral properties
similar to those of the prompt $\gamma$-ray emissions of GRBs detected so far. 
The low- and high-energy photon indices $\alpha_B$ and $\beta_B$ 
do not deviate significantly from $\simeq -1.0$ and $\simeq -2.5$, respectively 
(Figure~\ref{fig:ab}).
These values are quite typical for the prompt emissions of GRBs
\citep{preece00}.
Thus the prompt non-thermal emission may be categorized into X-ray Flashes
(XRFs), which are the transient events with properties similar to GRBs
except lower spectral peak energies and smaller fluences \citep[e.g.,][]{sakamoto05}.
The averaged spectral peak energy $\simeq 4.9$~keV and the isotropic-equivalent
energy $E_{\gamma,\iso} \simeq 6\times10^{49}$~erg of the non-thermal emission
obey the well-known Amati correlation, which is satisfied by the GRBs and XRFs
with known redshifts
except for outliers such as GRB 980425 and GRB 031203
\citep{amati02,amati06}.
Moreover, it has been reported that
this event obeys the lag-luminosity correlation \citep{liang06d,gehrels06}.

We may confirm that the speed of the emitting region was close to the light
speed.
For the emission radius $r_0$, the condition of the transparency for
the Compton scattering off the electrons associated with baryons is given 
in the non-relativistic formulation by
\begin{equation}
\tau \simeq \frac{\sigma_T E_{\iso}}{2\pi m_p c^2 {\beta_0}^2 {r_0}^2} < 1,
\end{equation}
where $E_{\iso}$ is the isotropic-equivalent kinetic energy of the outflow,
$\sigma_T$ is the Thomson cross section, and $m_p$ is the proton mass.
Here we assume that the outflow is dominated by the kinetic energy,
and $E_{\iso}$ is larger than the photon energy $E_{\gamma,\iso}$.
Estimating that $r_0 \sim c \beta_0 \delta t$ and adopting the values
$E_{\gamma,\iso} \simeq 6\times10^{49}$~erg and $\delta t\simeq 3000$~s,
we obtain $\beta_0 > 0.9$.
This corresponds to the Lorentz factor $\Gamma_0 > 2$.
For the relativistic wind, the condition of the transparency is given by 
$\tau \simeq \sigma_T L_{\iso}/(8\pi m_p c^3 {\Gamma_0}^3 r_0) < 1$
\citep{daigne02,meszaros06},
where $L_{\iso}$ is the isotropic-equivalent kinetic luminosity of the wind.
Estimating that $L_{\iso} > E_{\gamma,\iso}/\delta t$ and 
$r_0 \sim {\Gamma_0}^2 c \delta t$, we obtain no limit on the Lorentz 
factor.
(Since the maximum photon energy detected is lower than the electron's rest
energy $m_e c^2$, there is no limit on the Lorentz factor from the photon
annihilation or from the Compton scattering off pair-produced electrons 
and positrons \citep{kaneko06,lithwick01}.)
Therefore the expansion speed of the outflow should be close to the light speed,
although it is not necessarily ultra-relativistic.

All these facts show that the prompt non-thermal emission of this event
is a typical GRB (or XRF), except that its luminosity is extremely low and 
it is accompanied by the soft X-ray thermal component.
This is one of our motivations for considering that GRB 060218 arises from
a jet like typical GRBs in \S~\ref{sec:afterglow} and  \ref{sec:collapsar}.
In the rest part of this section, we focus on the decay phase of the prompt
non-thermal emission to suggest that its emission process does not stop abruptly
but decays slowly to connect to the detected anomalous X-ray afterglow.

\subsubsection{The decay phase}
\label{sec:decay}

The decay phase of the prompt non-thermal emission is well described by
$F(t) \propto t^{-2.0}$ and $E_p(t) \propto t^{-1.6}$, which implies a 
hardness-intensity correlation $F \propto E_p^{1.3}$.
Such a correlation, $F \propto {E_p}^{\zeta}$, can be seen in the decay phases of 
the GRB prompt emissions observed by BATSE, 
which has a broad distribution of the power-law indices $\zeta$, 
with $0.6 \lesssim \zeta \lesssim 3$ \citep{borgonovo01,ryde02}.
{\it Swift} satellite has succeeded in observing the decay phases of many GRB
prompt emissions more deeply than the observations in the pre-{\it Swift} era.
The result is that
the distribution of the temporal indices of the flux is 
also broad; $-5 \lesssim \alpha \lesssim -1$, where $F \propto t^{\alpha}$ 
\citep{obrien06}.
Thus the flux decay of the prompt non-thermal emission of GRB 060218 is 
relatively shallow.

First, we argue that the decay phase is not attributed to the kinematical effect due to
the curvature of the emitted region which ceased the emission process suddenly.
Since the emitting shell should have a curvature, the observer receives the radiation 
far from the line of sight after the cessation of the emission process.
The region at higher latitude from the line of sight has a lower velocity towards
the observer, so that the emission becomes dimmer and softer progressively because
of the relativistic beaming effect.
This effect is the so-called curvature effect, and have been widely studied for 
the decay phases of GRB prompt emissions
\citep{fenimore96,sari97,kumar00,ryde02,dermer04}.
{\it Swift} observations have suggested that the decay phases of GRB prompt emissions
with the temporal indices $\alpha < -2$ are due to the curvature effect
\citep{liang06a,zhang06,lazzati06,yama06,kobayashi06}.

Suppose that the shell moving with a bulk Lorentz factor $\Gamma_0$ 
ceases the emission at a radius $r_0$ and at a time $t_0$ and that the line of sight
is within the shell, i.e., the observer views the shell on-axis.
The comoving frequency $\nu'$ is boosted to 
$\nu = \nu'{\mathcal D}$ in the observer's frame,
where ${\mathcal D} = [\Gamma_0(1-\beta_0\cos\theta)]^{-1}$ is the Doppler factor.
Because the observed time is given by $t = t_0 - r_0\cos\theta/c$,
the Doppler factor is related to the observed time by
\begin{equation} 
{\mathcal D} = \frac{r_0}{c\beta_0\Gamma_0}(t-t_r)^{-1},
\end{equation}
where $t_r \equiv t_0 - r_0/(c\beta_0)$ is the departure time of the 
shell from the central engine.
Thus the time-resolved spectral peak energy evolves as
\begin{equation} 
E_p \propto {\mathcal D} \propto (t-t_r)^{-1}.
\end{equation}
The received flux in a given time interval $dt$ is
$F_{\nu} dt \propto j'_{\nu'} \mathcal{D}^2 d\Omega$,
where $j'_{\nu'}$ is the comoving surface brightness, and
the solid angle of the emitting region is related to the observed time 
interval by $d\Omega = 2\pi d(\cos\theta) \propto dt$ 
\citep{rybicki79,granot99,ioka01}.
Thus the received flux is estimated by
\begin{equation}
F_{\nu} \propto j'_{\nu'}{\mathcal D}^2 \propto (\nu')^{1+\beta_B}{\mathcal D}^2
\propto \nu^{1+\beta_B}{\mathcal D}^{1-\beta_B} \propto \nu^{1+\beta_B}(t-t_r)^{-1+\beta_B},
\end{equation} 
where we assume the high-energy Band spectrum $j'_{\nu'}\propto
\nu'^{1+\beta_B}$
because the $15-150$~keV band is above the peak energy $E_p$.
With the observed value $\beta_B \simeq -2.5$, we have
$F(t) \propto (t-t_r)^{-3.5}$ and $E_p(t) \propto (t-t_r)^{-1}$.
To reproduce the temporal index obtained from our analysis, $t_r < 0$ is required 
for $F(t)$, while in contrast, $t_r > 0$ is required for $E_p(t)$.
Therefore $F(t)$ and $E_p(t)$ cannot be fitted simultaneously by this
model. 
One can also consider the structured jet model in which the spectrum and brightness
vary in the angular direction \citep{rossi02,zhang02a,yama06,dyks06}.
However, the shallower flux decay requires a jet with brighter rim,
which seems implausible.

Thus the emission process is attributed not to the curvature effect of the shell which 
ceases the emission abruptly, but to slowly decaying emission
processes.
The first possibility of the decaying emission is 
the synchrotron emission from the external shock \citep{rees92,sari98}.
Since the high-energy range of the spectrum will be above the cooling frequency,
the high-energy photon index $\beta_B \simeq -2.5$ corresponds to the energy
distribution index of the accelerated electrons $p \simeq 3$.
Then the flux should decay as
$F(t) \propto t^{\frac{1}{2}-\frac{3}{4}p} \sim t^{-1.75}$.
The characteristic frequency $\nu_m$ evolves as $t^{-3/2}$.
These temporal behaviors are not inconsistent with the results of our analysis.
However, the standard model gives much lower 
characteristic frequency $\nu_m$ than the observed one
$\sim 10^{18}$~Hz at the deceleration time $t \simeq 500$~s
with the X-ray luminosity $L_X \sim 10^{47}~{\rm erg}~{\rm s}^{-1}$.
Therefore this model is not so appealing.
 
The second possibility is that the decaying emission is attributed to
the central engine activity.
The decaying flux in $15-150$~keV range, if extrapolated as $F(t) \propto t^{-2.0}$,
becomes $\sim 10^{-11}~{\rm erg}~{\rm s}^{-1}$ at $t \sim 10^4$~s.
This flux is comparable to that of the detected anomalous X-ray afterglow 
in $0.3-10$~keV range, decaying as $F_X \propto t^{-1.1}$ with the photon index 
$\beta_X \simeq -3.2$, which cannot be explained within the external shock
synchrotron model \citep{campana06,soderberg06b}.
If the anomalous
X-ray afterglow is a continuation of the prompt non-thermal emission
(i.e., the X-ray emission detected at $t \gtrsim 10^4$~s is not an afterglow but
the decaying prompt emission), the prompt non-thermal emission of this
burst is very long ($> 10^6$~s)
and it is necessary that the photon index varies from 
$\beta_B \simeq -2.5$ to $\beta_X \simeq -3.2$.
\citet{ghisellini06b} have also suggested such a scenario and 
shown that a synchrotron inverse-Compton model
could reproduce the UV/optical thermal component as well as 
the prompt non-thermal emission and the anomalous X-ray afterglow.
Here we note that the flux decays steeper than $t^{-1}$ 
after the peak
time $t \simeq 500$~s, 
so that the time-integrated radiation energy 
does not increase so much after the peak time. 
Thus the total radiation energy of the prompt non-thermal emission
remains $E_{\gamma,\iso} \simeq 6\times10^{49}$~erg.

Therefore the engine of this event could be active for at least $10^6$~s.
Recent detailed observations of GRB X-ray afterglows have revealed
that a number of bursts show erratic X-ray flares, which may be linked to
the restartings of the central engine after the prompt emission phase
\citep[e.g.,][]{zhang06,burrows05b,ioka05}.
Since some X-ray flares are discovered $\sim 10^5$~s after the trigger,
the engines of typical GRBs could be also active for $\gtrsim 10^5$~s.
However, they are spiky and intermittent activities, while this event shows
the smooth and power-law activity.
We speculate that this indicates that the engine of this event is different 
from those of typical GRBs which are believed to be black holes.
Although this is just a speculation, it agrees with the analysis of the 
SN component of this event, 
which suggests that the progenitor star was less massive and collapsed to
a neutron star instead of a black hole \citep{mazzali06}.
Further discussion will be presented in \S~\ref{sec:discussion}.

\section{A Jet Model of the Radio Afterglow}
\label{sec:afterglow}

The afterglow of this burst shows chromatic light curves.
The UV/optical afterglow is quite dim and dominated by the thermal component
and the SN component \citep{campana06,ghisellini06b}.
The X-ray afterglow has a spectrum much steeper than those of typical
GRB X-ray afterglows \citep{soderberg06b,butler06},
and it could be a continuation of the prompt non-thermal emission 
(see \S~\ref{sec:decay}).
Only the radio afterglow seems rather typical and can be explained 
by the standard external shock synchrotron model
\citep{soderberg06b,fan06}.
In this section, we show that the available radio data may arise
from an initially collimated outflow. 
In the spherical outflow model, it seems difficult for a SN shock to provide 
the relativistic spherical ejecta with sufficient energy, 
as stated in \S~\ref{sec:intro}.
In the jet model, taking into account the UV/optical and X-ray afterglows,
we will constrain the initial opening angle and Lorentz factor of the
jet as $\theta_0 \simeq 0.3$ and $\Gamma_0 \simeq 5$, respectively.
We also show that the off-axis scenario in the jet model is less likely 
because it requires higher $\gamma$-ray efficiency than the on-axis scenario.

\subsection{The radio afterglow}
\label{subsec:radio}

For $2-22$~days the afterglow flux at $8.46$~GHz is well detected
and shows the power-law decay as $t^{-0.85}$.
The spectrum at $t \simeq 5$~days has a peak between $1.43$~GHz and $4.86$~GHz,
and this peak can be interpreted as the self-absorption frequency
$\nu_a$ in the standard model.
This peak would not be the typical synchrotron frequency $\nu_m$ because
of the small $1.43$~GHz flux.
Thus the $8.46$~GHz is most likely in the frequency range 
$\nu_m < \nu_a < \nu < \nu_c$,
where $\nu_c$ is the cooling frequency.

In the standard model, the temporal index of the flux at each frequency
range can be calculated for several cases;
for a relativistic/non-relativistic blastwave expanding into circumburst
medium of constant/wind-profile density, or for a sideways expanding 
relativistic blastwave.
The temporal index of the flux at the frequency range
$\nu_m < \nu_a < \nu < \nu_c$ for each case is 
$-\frac{3}{4} p + \frac{3}{4}$ (relativistic blastwave, constant density), 
$-\frac{3}{2} p + \frac{21}{10}$ (non-relativistic blastwave, constant density),
$-\frac{3}{4} p + \frac{1}{4}$ (relativistic blastwave, wind-profile density), 
$-\frac{7}{6} p + \frac{5}{6}$ (non-relativistic blastwave, wind-profile density), 
and $-p$ (sideways expanding),
where $p$ is the index of the energy distribution function of the 
accelerated electrons
\citep[e.g.,][]{livio00,sari98,sari99,chevalier99,frail00}.
For each case to reproduce the observed temporal index $-0.85$, 
$p$ is required to be $\simeq 2.1, 2.0, 1.5, 1.4,$ and $0.9$, respectively.
The values $p \gtrsim 2$ are typical for the GRB afterglows
\citep[e.g.,][]{panaitescu02,yost03}, and for this reason, we focus on the 
first two possibilities (i.e., the constant density cases).
Note, however, that the possibilities $p<2$ (i.e., the wind-profile density
cases and the sideways expanding case) cannot be ruled out.

For the two possibilities of the standard model,
the available radio data requires the following three conditions:
\begin{itemize}
\item[(a)] The spectrum peaks at $\nu_a \sim 4\times10^9$~Hz at $t\simeq 5$~days.
\item[(b)] The flux at $22.5~\rm{GHz}$ is $\sim 0.25$~mJy at $t\simeq3$~days.
\item[(c)] The cooling frequency is $\nu_c \leq 5\times10^{15}$~Hz so that the
 synchrotron spectrum of the external shock electrons does not dominate the detected 
anomalous X-ray afterglow.
\end{itemize}

\subsubsection{Relativistic blastwave model}
\label{sec:rela}

\citet{fan06} have modeled the first possibility, i.e.,
the synchrotron spectrum from the relativistic blastwave 
with $p \simeq 2.1$ expanding into the constant density medium, and derived three
constraints on the physical parameters of the afterglow from the above three conditions
$(a)-(c)$ (their Equations $(8)-(10)$).
The satisfying parameters they have suggested are the isotropic-equivalent
kinetic energy of the outflow $E_{k,\iso} \sim 10^{50}$~erg, the number density
of the circumburst medium $n \sim 10^2~{\rm cm}^{-3}$, and the ratios of the 
accelerated electrons energy and the magnetic energy to the shocked thermal energy
$\epsilon_e \sim 10^{-2}$ and $\epsilon_B \sim 10^{-3}$, respectively.
With these parameters, the Lorentz factor of the outflow is estimated as
$\Gamma \approx 2~(t/1~{\rm day})^{-3/8}$.
If the opening angle of the outflow is $\theta_0 \leq 1$, the outflow will
begin to expand sideways at $t \leq 6$ days and the radio light curves will break
\citep{rhoads99}.
However, the actual light curve does not show such a break, so that it is 
concluded that $\theta_0 > 1$ in this model, i.e., the outflow is spherical.
In this case, the outflow becomes non-relativistic at $t \simeq 6$ days, 
and the radio flux varies from $t^{-0.85}$ to $t^{-1.1}$. 
This is consistent with
the available radio light curves \citep{fan06}.

The parameters $E_{k,\iso} \sim 10^{48}$~erg, $n \sim 10^2~{\rm cm}^{-3}$,
$\epsilon_e \sim 10^{-1}$, and $\epsilon_B \sim 10^{-1}$ also satisfy the 
above three constraints and fit the radio data \citep{soderberg06b}.
However, since the isotropic-equivalent $\gamma$-ray energy is 
$E_{\gamma,\iso} \simeq 6\times10^{49}$~erg, the inferred $\gamma$-ray 
efficiency is extremely high;
$\eta_{\gamma} \equiv E_{\gamma,\iso}/(E_{\gamma,\iso} + E_{k,\iso}) \sim
98\%$, 
and thus this parameter set is implausible.

The relativistic spherical outflow is not consistent with the standard 
collapsar scenario, in which a relativistic collimated outflow digs 
a narrow hole in the progenitor star without loading so much stellar matter
\citep{woosley93,macfadyen99}.
Thus one may consider a scenario
in which a spherical outflow is accelerated to a mildly relativistic
speed as the SN shock sweeps the 
stellar envelope with steep density gradient \citep{matzner99,tan01}.
However, it seems impossible that the required kinetic energy 
$E_{k,\iso} \sim 10^{50}$~erg is transferred into the relativistic ejecta 
by the SN with the total kinetic energy $\simeq 10^{51}$~erg \citep{mazzali06}, 
as discussed in \S~\ref{sec:intro}.
For this reason, we conclude that the first possibility is quite unlikely.

\subsubsection{Non-relativistic blastwave model}
\label{sec:non-rela}

Now we investigate the second possibility; the radio afterglow may arise from
the non-relativistic phase of an initially jetted outflow with $p \simeq 2.0$ 
expanding into the constant density medium.
The relativistic jet is decelerated and subsequently expands sideways,
and finally becomes non-relativistic at a certain transition time $t_s$.
The transition time is estimated by requiring that the speed of the Sedov-Taylor
blastwave is close to the light speed, i.e., 
$\beta = \frac{2}{5c}[E_k/(n m_p t^3)]^{1/5} \sim 1$,
where $E_k$ is the kinetic energy of the spherical blastwave \citep[see][]{livio00}:
\begin{equation}
t_s \simeq 7.5 \times 10^6 ~{\rm s} 
\left(\frac{E_{k,51}}{n}\right)^{1/3}.
\label{eq:sedov}
\end{equation}
Here (and hereafter) we have adopted the notation $Q = 10^x Q_x$ in cgs units.
Since the available $8.46$~GHz lightcurve does not show the sideways expanding
relativistic phase $t^{-p} \sim t^{-2}$, we require $t_s \lesssim 2$~days.

Three constraints on the physical parameters of the afterglow are given by
the conditions $(a)-(c)$ in
the first part of this subsection (\S~\ref{subsec:radio}).
Assuming that the minimum Lorentz factor of the accelerated electrons is
$\gamma_m \simeq 10^{-1} \beta^2 \epsilon_e m_p/m_e$,
we may derive the following three constraints on the physical parameters:
\begin{eqnarray}
{\epsilon_{e,-1}}^{1/3} {\epsilon_{B,-2}}^{1/3} {E_{k,51}}^{1/3}
n^{1/3} &\sim 1, \label{eq:c1} \\
{\epsilon_{e,-1}} {\epsilon_{B,-2}}^{0.75} {E_{k,51}}^{1.3}
n^{0.45} &\sim 3\times10^{-3}, \label{eq:c2} \\
{\epsilon_{B,-2}}^{1.5} {E_{k,51}}^{0.6} n^{0.9} &\geq 0.4.
\label{eq:c3}
\end{eqnarray}
The results are similar to those derived in the relativistic blastwave model 
\citep[see \S~\ref{sec:rela};][]{fan06}.
These constraints and the requirement of $t_s \lesssim 2\times10^5$~s are
satisfied by
$E_k \simeq 2\times10^{48}~{\rm erg}, n \simeq 10^2~\rm{cm}^{-3},
\epsilon_e \simeq 10^{-1},$ and $\epsilon_B \simeq 10^{-1}$.

Then the isotropic-equivalent kinetic energy before sideways expansion is estimated
by $E_{k,\iso} = 2E_k/{\theta_0}^2 
\simeq 4\times10^{48}~{\theta_0}^{-2}~{\rm erg}$.
To obtain a reasonable $\gamma$-ray efficiency 
$\eta_{\gamma} = E_{\gamma,\iso}/(E_{\gamma,\iso} + E_{k,\iso}) \lesssim 0.5$
\citep{lloyd04,ioka06,fanpiran06,granot06,zhang06b},
the opening angle $\theta_0 \lesssim 0.3$ is favorable.
In the following section, we will further constrain
the opening angle and also the Lorentz factor of the jet
using the observed UV/optical and X-ray emission.

\subsection{Constraints from the UV/optical and X-ray afterglow}

In our model ($E_k \simeq 2\times10^{48}~{\rm erg}, n \simeq 10^2~\rm{cm}^{-3},
\epsilon_e \simeq 10^{-1},$ and $\epsilon_B \simeq 10^{-1}$),
the characteristic parameters at $t \simeq 5$~days are calculated as
$F_{\nu,\rm{max}} \sim 2\times10^{-25}~\rm{erg}~\rm{cm}^{-2}~\rm{s}^{-1}
~\rm{Hz}^{-1}$, $\nu_m \simeq 5\times10^6$~Hz, $\nu_a \simeq 4\times10^9$~Hz,
and $\nu_c \simeq 7\times10^{13}$~Hz.
Thus the optical $\nu F_{\nu}$ flux at this time is  
$\sim 10^{15} F_{\nu,\rm{max}} (\nu_c/\nu_m)^{-0.5} (10^{15}/\nu_c)^{-1}
\simeq 4\times10^{-15}~\rm{erg}~\rm{cm}^{-2}~\rm{s}^{-1}$.
If we trace back to earlier times through the standard jet model,
the optical flux gets much higher than this value
(see Figure~\ref{fig:optmodel}).
The detected optical flux at $10^4-10^5$~s is dominated by the thermal component with 
$\nu F_{\nu} \sim 10^{-11}~\rm{erg}~\rm{cm}^{-2}~\rm{s}^{-1}$
\citep{ghisellini06b,campana06}.
The requirement that the synchrotron flux from our jet should not exceed this flux
gives constraints on the initial opening angle $\theta_0$ and 
the initial Lorentz factor $\Gamma_0$ of the jet.

The external shock of the jet progressively experiences 
the Blandford-McKee evolution, the sideways expanding evolution, 
and the Sedov-Taylor evolution.
For each evolution phase the characteristic frequencies of the synchrotron 
spectrum with $p \simeq 2.0$
vary as $\nu_m \propto t^{-1.5}, t^{-2},$ and $t^{-3}$; 
$\nu_a (>\nu_m) \propto t^{-2/3}, t^{-0.5},$ and $t^{-2/3}$;
$\nu_a (<\nu_m) \propto t^{0}, t^{-0.2},$ and $t^{1.2}$;
and $\nu_c \propto t^{-0.5}, t^{0},$ and $t^{-0.2}$
\citep[e.g.,][]{sari98,sari99,livio00}.
If we trace back to $t \sim 10^4$~s, $\nu_m, \nu_a,$ and $\nu_c$
all remain lower than the optical band.
Thus for $t \gtrsim 10^4$~s, the optical band is at the range 
$\nu > \nu_c$, and the optical energy flux evolves as
$t^{-1}$ at the Blandford-McKee phase, 
$t^{-2}$ at the sideways expanding phase, and
$t^{-1}$ at the Sedov-Taylor phase.
With these behaviors of the optical flux from the external shock,
we can suggest a following possible jet scenario
(see Figure~\ref{fig:optmodel} for
a schematic picture of the optical lightcurve).
The external shock of the jet is decelerated at 
$t_{\dec} \simeq 6\times10^3$~s,
begins to expand sideways at $t_j \simeq 2\times10^4$~s, and
shifts to the Sedov-Taylor expansion at $t_s \simeq 2\times10^5$~s.
The peak optical flux is $\nu F_{\nu} \simeq 3\times10^{-12}$, 
which is less than the observed one, so that
the optical afterglow estimated within our jet model is not inconsistent
with the available optical data.

Since the X-ray band is also above $\nu_c$ and $p \simeq 2.0$,
the X-ray $\nu F_{\nu}$ flux is comparable to the optical one.
Thus this external shock emission does not overwhelm the observed anomalous
X-ray afterglow shown in \citet{campana06} and \citet{soderberg06b}.

Now we give the reasonable values of $\theta_0$ and $\Gamma_0$.
At the sideways expansion phase, the opening angle of the jet
evolves as $\theta = \Gamma^{-1} \propto t^{1/2}$ \citep{rhoads99,sari99}, and finally
becomes $\theta \simeq 1$ at the transition time $t_s$ to the Sedov-Taylor
expansion phase.
Thus the initial opening angle of the jet is determined by
\begin{equation}
\theta_0 \left(\frac{t_s}{t_j}\right)^{1/2} \simeq 1.
\end{equation}
When $t_s \simeq 2\times10^5$~s and $t_j \simeq 2\times10^4$~s are
adopted, we obtain $\theta_0 \simeq 0.3$.
If $\theta_0$ is smaller, the ratio $t_s/t_j$ is larger.
At the Blandford-McKee phase, the Lorentz factor of the jet
evolves as $\Gamma \propto t^{-3/8}$, and finally becomes 
$\Gamma \simeq {\theta_0}^{-1}$ at the jet-break time $t_j$.
Thus the initial Lorentz factor of the jet is determined by
\begin{equation}
\Gamma_0 \left(\frac{t_j}{t_{\dec}}\right)^{-3/8} \simeq 
{\theta_0}^{-1}.
\end{equation}
When $t_j \simeq 2\times10^4$~s and $t_{\dec} \simeq 6\times10^3$~s
are adopted, we obtain $\Gamma_0 \simeq 5$.
If $\Gamma_0$ is larger, the ratio $t_j/t_{\dec}$ is larger.
The upper bound of the optical data does not allow the values of 
$t_s/t_j$ and $t_j/t_{\dec}$ larger than those we have adopted, and 
therefore it does not allow the smaller $\theta_0$ and the larger $\Gamma_0$.
Since the consideration about the $\gamma$-ray efficiency requires
$\theta_0 \lesssim 0.3$ (see \S~\ref{sec:non-rela}), 
the opening angle is restricted to 
a narrow range around $\theta_0 \simeq 0.3$.

In a summary, we have obtained a jet model for the radio afterglow of GRB 060218,
with $\theta_0 \simeq 0.3$, $\Gamma_0 \simeq 5$, $E_{k,\iso} \simeq 4\times10^{49}$~erg,
and $\eta_{\gamma} \simeq 0.6$, which is compatible with the UV/optical and 
X-ray data.
The collimation-corrected energy of the jet is 
$E_j \simeq (E_{\gamma,\iso} + E_{k,\iso}) {\theta_0}^2/4 \simeq 10^{48}$~erg.
Assuming that the duration of the prompt non-thermal emission $\delta t \sim 10^3$~s
is the active time of the central engine $\delta T$, the luminosity 
of the jet is estimated by $L_j \sim E_j/\delta T \sim 10^{45}~{\rm erg}~{\rm s}^{-1}$.

\subsection{Off-axis scenario}

In the above arguments in \S~\ref{sec:prompt} and \ref{sec:afterglow}, 
we have assumed that the observer views the jet from the on-axis direction 
and the luminosity of the prompt non-thermal emission is intrinsically low.
It is possible that the low luminosity of GRB 060218 is 
attributed to a typical bright GRB jet viewed off-axis 
\citep[e.g.,][]{yama03,ramirez05,granot05,toma05}.
However, we find that the off-axis scenario is less likely than the on-axis 
scenario since higher $\gamma$-ray efficiency is required.

Since the outflow is spherical when the radio afterglow is observed,
the off-axis scenario is the same as the on-axis one in this phase.
Thus the isotropic-equivalent kinetic energy is determined
by $E_{k,\iso} = 2E_k/{\theta_0}^2 
\simeq 4\times10^{48}~{\theta_0}^{-2}$~erg.
In the off-axis scenario, the isotropic-equivalent $\gamma$-ray energy
$E_{\gamma,\iso,{\rm on}}$ 
measured if viewed on-axis is larger than the observed one
$E_{\gamma,\iso} \simeq 6\times10^{49}$~erg.
So the $\gamma$-ray efficiency is expected to be higher than 
that estimated within the on-axis scenario \citep[e.g.,][]{toma06}.

To obtain the smaller $\gamma$-ray efficiency, we could consider $\theta_0 < 0.3$.
However, the smaller $\theta_0$ leads to even higher $\gamma$-ray efficiency,
as we will see below.
The smaller $\theta_0$ requires $t_j < 10^4$~s, and then 
the optical flux, if received from the on-axis direction, exceeds the constraint
from the detected optical afterglow (see Figure~\ref{fig:optmodel}).
To escape the constraint, the synchrotron flux from the off-axis jet should peak
at some time $t_v$ after $t_j$.
The peak occurs when the line of sight enters the relativistic beaming cone of 
the emission from the edge of the sideways expanding jet, i.e., 
$\theta_v \sim \theta + {\Gamma}^{-1} \simeq 2\Gamma^{-1}$.
The Lorentz factor varies as $\Gamma \propto t^{-1/2}$ in the sideways expansion phase
and becomes $\Gamma \sim 1$ at $t_s$, and thus the following equation is satisfied:
\begin{equation}
\frac{1}{2} \theta_v \left(\frac{t_s}{t_v}\right)^{1/2} \simeq 1.
\end{equation}
Since $t_v$ should be greater than $10^4$~s, we obtain $\theta_v \gtrsim 0.4$.
The isotropic-equivalent $\gamma$-ray energy measured if viewed on-axis is 
roughly estimated by 
$E_{\gamma,\iso,{\rm on}}
\approx [\Gamma_0(\theta_v-\theta_0)]^6 E_{\gamma,\iso}
> 10^{47}~{\Gamma_0}^6$~erg,
where we have used $\theta_v > \theta_0$.
With $\Gamma_0 \geq {\theta_0}^{-1}$, we obtain 
$E_{\gamma,\iso,{\rm on}} > 10^{47}~{\theta_0}^{-6}$~erg.
For $\theta_0 < 0.3$, this leads to higher $\gamma$-ray
efficiency than the on-axis scenario; $\eta_{\gamma} > 0.7$.
Therefore we may conclude that the off-axis scenario is unlikely for this event.

\section{A Collapsar Model of GRB 060218}
\label{sec:collapsar}

In the previous section, we have obtained a possible jet model 
that is consistent with the available afterglow data.
Our jet model requires
an opening angle $\theta_0 \simeq 0.3$, which is somewhat larger than
those of typical cosmological GRBs \citep[e.g., see][]{ghirlanda04}, 
and an extremely small luminosity $L_j \sim 10^{45}~{\rm erg}~{\rm s}^{-1}$, 
which is about $10^5$ times smaller than those of typical cosmological
GRBs.
Then one may wonder
whether such a wide and low-luminosity jet can penetrate a progenitor star.
It is possible that a jet becomes non-relativistic if it loads
so much stellar matter with a large cross section.
\citet{matzner03} has discussed analytically 
several conditions for making a hole in
the collapsar model.
We extend his theoretical considerations to conclude that 
an adiabatic cold jet is excluded, but a non-adiabatic hot jet is appropriate
for this event.
(Our argument can be also applied to typical GRB jets,
and some wide GRB jets also would need to be hot.)

\subsection{The motion of a jet head}

The standard collapsar model assumes that a black hole
or a neutron star with an accretion disk is formed after an iron core of the 
massive progenitor star collapses and that the system produces a jet 
\citep{woosley93,macfadyen99}.
Since the free-fall timescale of a stellar envelope is longer than, or comparable with,
the crossing timescale for the jet to propagate 
within the progenitor star and to hit its
surface, the stellar envelope would disturb the progress of the jet.

Consider the progress of a relativistic jet outward through the stellar envelope.
Two shocks form: a reverse shock reducing the jet speed and increasing its internal 
energy; and a forward shock that propagates into the surrounding stellar envelope
giving it internal energy.
As a result, there are four distinct regions in this system: the propagating jet;
the head of the jet lying between the two shocks; the stellar envelope; and 
a cocoon consisting of shocked jet and shocked ambient material
\citep[see Figure~2 of][]{matzner03}.
This system is similar to classical double radio sources \citep{begelman89}.

Two conditions are required for a jet to break out the star relativistically.
First, the Lorentz factor of the jet head $\Gamma_h$ should be smaller than the inverse
of the opening angle of the jet $\theta$.
If so, the shocked material may escape sideways to form the cocoon, and the jet may 
avoid baryon loading.
Second, the speed of the jet head should be larger than that of the expanding cocoon,
\begin{equation}
\beta_h > \beta_c.
\label{eq:jetdriven}
\end{equation}
If this condition is violated, the cocoon expands spherically around the jet and 
finally explode the star, producing a non-relativistic dense spherical outflow.
When the above two conditions are satisfied, we may consider the longitudinal balance
of the momentum flux in the frame of the jet head,
\begin{equation} 
w_j {\Gamma_j}^2 {\Gamma_h}^2 (\beta_j - \beta_h)^2 + p_j 
= w_a {\Gamma_h}^2{\beta_h}^2 + p_a.
\label{eq:longitudinal}
\end{equation}
Here the subscripts $j$ and $h$ describe the jet and the jet head, respectively,
and $w (\equiv e + p)$, $p$, and $e$ are the enthalpy, the pressure, and the energy density
including rest energy density $\rho c^2$, respectively.
For stellar envelopes, $p_a \ll \rho_a c^2$ and $w_a = \rho_a c^2$ are good approximations.

If $\Gamma_j$ is significantly larger than unity, $p_j$ can be neglected
in the left-hand side of Equation~(\ref{eq:longitudinal}), and we obtain
$({\beta_h}^{-1} -1)^{-2} = w_j {\Gamma_j}^2 / (\rho_a c^2) 
= L_{\iso}/(4 \pi r^2 \rho_a c^3)$, where at the last equality we introduce
the isotropic-equivalent luminosity $L_{\iso}$ of the jet.
For the jet to drive the explosion of the star, this equation should be satisfied
when $r = R$, where $R$ is the radius of the progenitor star, so that 
$({\beta_h}^{-1} -1)^{-2} \approx 10^{-3} L_{\iso,51}
(R/R_{\odot})^{-2} {\rho_a}^{-1}$.
This factor is much less than unity for the parameters of typical GRBs
(and so for the low-luminosity GRBs),
so that the jet head is found to be non-relativistic.
Thus, with the collimation-corrected luminosity of the jet 
$L_j = w_j {\Gamma_j}^2 c \pi r^2 {\theta}^2$, 
we obtain
\begin{equation}
\beta_h \approx \left(\frac{L_j}{\pi r^2 \theta^2 \rho_a c^3}\right)^{1/2}.
\label{eq:beta_h}
\end{equation}
We take $\rho_a \sim M/(4\pi R^3/3)$ for simplicity, where $M$ is the total mass
of the progenitor star.

\subsection{Cocoon structure}

The cocoon drives a shock into the stellar envelope, which
expands non-relativistically at the velocity $\beta_c$ given by
the transverse balance of the cocoon pressure and the ram pressure of the stellar
envelope,
\begin{equation}
p_c = \rho_a c^2 {\beta_c}^2.
\label{eq:cocoon_ambient}
\end{equation}
The pressure of the cocoon $p_c$ is roughly constant away from the jet head,
and is approximated by $E_{\rm in}/(3V_c)$ with the thermal energy $E_{\rm in}$
deposited in the cocoon and the cocoon volume $V_c$.
Since the jet head is non-relativistic, most of the kinetic energy of the jet
gets thermalized through the reverse shock.
The cocoon energy is the energy caught by the jet head, which is calculated by
$E_{\rm in} = \int^r_0 L_j \frac{dr}{c\beta_h} \simeq L_j \frac{r}{c\beta_h}$.
The cocoon length is given by $r \simeq c\beta_h t$ as long as 
$\beta_h > \beta_c$ is satisfied,
and the cocoon width is given by $R_c \simeq c\beta_c t$.
So the cocoon volume is roughly $V_c \simeq (\pi/3) {R_c}^2 r 
\simeq (\pi/3)r^3 {\beta_c}^2/{\beta_h}^2$.
Therefore the cocoon pressure can be written by 
$p_c \simeq L_j\beta_h/(\pi r^2 c {\beta_c}^2)$.
Equation~(\ref{eq:cocoon_ambient}) gives us the following equations:
\begin{eqnarray}
\beta_c = \left(\frac{L_j}{\pi r^2 \rho_a c^3} \beta_h \right)^{1/4},
\label{eq:beta_c} \\
p_c = \left(\frac{L_j}{\pi r^2} \rho_a c \beta_h \right)^{1/2},
\label{eq:p_c}
\end{eqnarray}  
which describe the velocity and the pressure of the cocoon with the parameters 
of the jet and the stellar envelope within the assumption of $\beta_h > \beta_c$.

\subsection{Cold jet}

Here we examine an adiabatic jet that propagates ballistically in the progenitor star
\citep[see also][]{meszaros01,waxman03}.
An important point is that
the opening angle and the Lorentz factor of the cold ballistic jet do not change
after exiting the progenitor star.
If the opening angle and Lorentz factor of the jet at the breakout are
$\theta_{\br}$ and $\Gamma_{\br}$, 
they are also
$\theta_0 = \theta_{\br}$ and $\Gamma_0 = \Gamma_{\br}$
after exiting the star.
In contrast, a hot jet which propagates non-adiabatically by the interaction
between the jet and the cocoon and keeps dominated by the internal energy changes
its opening angle and Lorentz factor through the free expansion
outside the star.
The hot jet will be discussed in the next subsection.

For the jet to drive an explosion, the consistency $\beta_h > \beta_c$
should be satisfied. From Equations~(\ref{eq:beta_h}) and
(\ref{eq:beta_c}), this leads to the constraint on the opening angle of the jet,
\begin{equation}
\theta < \left(\frac{L_j}{\pi r^2 \rho_{a} c^3}\right)^{1/6}.
\label{eq:consistency}
\end{equation}

For GRB 060218, the underlying SN is Type Ic, which implies that the progenitor
star is a C/O Wolf-Rayet star \citep[e.g.,][]{mazzali06}.
Its radius and mass are typically $R \sim 10^{11}$~cm and 
$M \sim 10^{34}$~g, respectively.
Then the averaged mass density $\rho_a$ is $\sim 1~{\rm g}~{\rm cm}^{-3}$.
For the luminosity $L_j \sim 10^{45}~{\rm erg}~{\rm s}^{-1}$, the opening angle of
the jet at the breakout (i.e., when $r = R$) is constrained to $\theta_{\br} < 0.03$,
so that $\theta_0 < 0.03$ within the cold jet scenario.
The value of the opening angle suggested in our jet model ($\theta_0 \simeq 0.3$)
violates this constraint. 
Therefore the cold jet scenario is excluded for this event.

A wide cold jet may be allowed if the jet is being launched long after
the cocoon explodes the star, since the ambient density $\rho_a$ 
drops through the expansion in Equation~(\ref{eq:consistency}).
However the star must expand from $R_c \sim 10^{11}$~cm to $\sim
10^{17}$~cm in order that $\theta_0 \simeq 0.3$ is allowed,
which is unlikely.

\subsection{Hot jet}

Next we consider 
the possibility that the jet is dominated by the internal energy
throughout the propagation in the progenitor star
\citep[see also][]{lazzati05}.
At the jet-cocoon boundary, oblique shocks and shear instabilities 
may occur and dissipate the kinetic energy of the jet.
This situation can be seen in several numerical simulations of the 
collapsar model \citep{aloy00,wzhang03,wzhang04,umeda05,mizuta06}.

In contrast to the cold jet, the hot jet expands freely after exiting the star.
As a result, the opening angle is determined by the Lorentz factor $\Gamma_{j,\ex}$ just
before exiting the star.
In the comoving frame of the hot material moving outward relativistically with the 
Lorentz factor $\Gamma_{j,\ex}$, the hot material will expand freely and finally
get the Lorentz factor $\Gamma' = e/{\rho_j c^2}$.
In the laboratory frame, the material will be beamed into the opening angle 
\begin{equation}
\theta_0 \sim {\Gamma_{j,\ex}}^{-1},
\label{eq:free}
\end{equation} 
and the Lorentz factor of the material is given by
$\Gamma_0 \sim \Gamma' \Gamma_{j,\ex} (1+\beta'\beta_{j,\ex}) 
\sim 2\Gamma'\Gamma_{j,\ex}$.
Note that $\Gamma_0 \theta_0 \sim 2\Gamma' > 1$ is satisfied.

For GRB 060218,
$\Gamma_{j,\ex} \sim \theta_0^{-1} \simeq 3$ is 
required to reproduce $\theta_0 \simeq 0.3$.
For the initial Lorentz factor of the jet exiting the star to be $\Gamma_0 \simeq 5$,
$\Gamma' = e/\rho_j c^2 \sim 1-2$ is favorable.
Therefore the jet of this burst may originate from a mildly hot jet 
in which the internal energy $e$ is comparable with the rest energy $\rho_j c^2$
before exiting the star.

\subsubsection{The breakout timescale}

Let us estimate the breakout timescale.
We expect that the breakout timescale is not much larger than
the active time of the central engine after the breakout $\delta T \sim 10^3$~s.
Otherwise, the engine has to stop suddenly just after breaking out the star,
and this seems unlikely.

For the hot jet, $p_j \gg \rho_j c^2$ and $w_j = 4p_j$ are good approximations,
and thus $p_j = L_j/(4{\Gamma_j}^2 c \pi r^2 \theta^2)$.
Before the breakout, the opening angle of the hot jet is determined by
the transverse pressure balance between the jet and the cocoon:
\begin{equation}
p_j = p_c.
\end{equation}
Using Equations~(\ref{eq:beta_h}) and (\ref{eq:p_c}), we obtain 
\begin{equation}
\theta = \left(\frac{L_j}{256\pi r^2 \rho_a c^3}\right)^{1/6} {\Gamma_j}^{-4/3}, 
\label{eq:hot_theta}
\end{equation}
which satisfies the consistency of $\beta_h > \beta_c$ (Equation~(\ref{eq:consistency})).
For GRB 060218, adopting the values $L_j \sim 10^{45}~{\rm erg}~{\rm s}^{-1}$, 
$R \sim 10^{11}$~cm, $\rho_a \sim 1~{\rm g}~{\rm cm}^{-3}$, and $\Gamma_j \simeq 3$,
we obtain the opening angle at the breakout time as 
$\theta_{\br} \sim 3\times10^{-3}$.

The velocities of the jet head (Equation~(\ref{eq:beta_h})) and the cocoon 
(Equation~(\ref{eq:beta_c})) are calculated by using Equation~(\ref{eq:hot_theta}),
\begin{eqnarray}
\beta_h = \left(\frac{16L_j}{\pi r^2 \rho_a c^3}\right)^{1/3} {\Gamma_j}^{4/3}, 
\label{eq:hot_beta_h} \\
\beta_c = \left(\frac{2L_j}{\pi r^2 \rho_a c^3}\right)^{1/3} {\Gamma_j}^{1/3}.
\label{eq:hot_beta_c}
\end{eqnarray}
The ratio of the two velocities is $\beta_h/\beta_c = 2\Gamma_j$.
Then, if $\Gamma_j$ is close to unity, we have $\beta_c \sim \beta_h$
and hence the cocoon expands quasi-spherically.
This situation is in a good agreement with recent series of numerical simulations
performed by \citet{mizuta06}.
They have argued that the morphology of the explosion depends on the Lorentz factor 
$\Gamma_{j,0}$ of the jet given at the inner boundary: 
when the Lorentz factor is high ($\Gamma_{j,0} \gtrsim 3$), the high pressure cocoon 
collimates the outflow to form a narrow, relativistic jet; 
and when the Lorentz factor is low, on the contrary, the outflow is not collimated
and expands quasi-spherically.

The breakout timescale is determined by $t_{\br} = R / (c\beta_{h,\br})$, where
$\beta_{h,\br}$ is the velocity of the jet head
(Equation~(\ref{eq:hot_beta_h})) obtained at $r=R$:
\begin{equation}
t_{\br} 
\simeq \left(\frac{16L_j}{\pi \rho_a}\right)^{-1/3} R^{5/3} {\Gamma_j}^{-4/3}.
\end{equation}
For GRB 060218, we obtain $t_{\br} \sim 300$~s, which is a reasonable value.
The hot jet of this event takes $t_{\br} \sim 300$~s to break out the progenitor star
and lasts further $\delta T \sim 10^3$~s to eject the material 
producing the prompt emission.

\subsubsection{A possible additional widening effect}

After the breakout time,
the opening angle of the jet before exiting the star
might become wider than the value $\theta_{\br} \sim 3\times10^{-3}$, which is 
determined by the pressure balance between the jet and cocoon.
The cocoon expands freely outward from the star after the breakout time, and finally
the cocoon pressure becomes negligible in $R/(c/\sqrt{3}) \sim 10$~s.
The cocoon shock that will sweep the stellar envelope is not so strong,
and the envelope would remain cold.
After the cocoon disappears, the jet opening angle would be determined by 
the transverse balance between 
the jet pressure and the ram pressure of the stellar envelope matter.
The balance equation in the comoving frame of the propagating jet is 
\begin{equation}
p_j = (\Gamma_j \rho_a)(r \dot{\theta})^2,
\end{equation}
where $r$ is the radial coordinate in the laboratory frame.
A Lorentz factor appearing in the right-hand side is due to the Lorentz transformation.
Substituting $p_j = L_j/(4{\Gamma_j}^2 \pi r^2 {\theta}^2 c)$, we find that $\theta$
gradually increases since the jet pushes the stellar envelope.
Let $L_j$ and $\Gamma_j$ be constant and integrate over time, and we obtain
\begin{equation}
\theta = \left(\frac{L_j}{\pi \rho_a c}\right)^{1/4} r^{-1} {\Gamma_j}^{-3/4} 
\delta T^{1/2}.
\end{equation}
For GRB 060218, if we adopt the values $L_j \sim 10^{45}~{\rm erg}~{\rm s}^{-1}$, 
$\rho_a \sim 1~{\rm g}~{\rm cm}^{-3}$, $r = R \sim 10^{11}$~cm, $\Gamma_j \simeq 3$, 
and $\delta T \sim 10^3$~s, we obtain $\theta \sim 0.1$.
This is less than the opening angle after free expansion $\theta_0
\simeq 0.3$ in Equation (\ref{eq:free}), 
so that the widening effect is not important in this event.
In some cases the widening effect could dominate the free expansion effect.

\section{Summary and Discussion}
\label{sec:discussion}

We have investigated whether GRB 060218 arises from a collimated jet.
So far the lack of the jet break
has led to the interpretation that the outflow of this event is
spherical, and thereby the outflow is not the standard collapsar jet
but the outermost parts of the stellar envelope that the SN shock accelerates to a 
mildly relativistic speed \citep{soderberg06b,fan06}.
However, we have shown that the available radio data may be interpreted as a 
non-relativistic phase of an initially collimated outflow within the standard external 
shock synchrotron model, and that the jet model with an initial opening angle  
$\theta_0 \simeq 0.3$, Lorentz factor $\Gamma_0 \simeq 5$, 
and a collimation-corrected luminosity $L_j \sim 10^{45}$~erg 
can explain the radio data and is compatible with the UV/optical and X-ray data.
This model is more natural than the initially spherical outflow model, because 
in the latter model, the relativistic ejecta for the 
prompt and afterglow emission with $\simeq 10^{50}$~erg could not be produced by
the underlying SN of the total kinetic energy $\simeq 10^{51}$~erg
\citep{mazzali06}.
Furthermore, the jet model is supported by the recent report of the detection of
optical linear polarization in the SN component of this event \citep{gorosabel06}.
We also show that the jet of this event can penetrate the progenitor star
by extending the analytical considerations by \citet{matzner03}.
The jet would be relativistically hot in the progenitor star and hence expand freely
outside the star into the relatively wide opening angle.
The off-axis scenario within the jet model is unlikely because it requires
an unrealistically high $\gamma$-ray efficiency.

With the jet opening angle $\theta_0 \simeq 0.3$, the collimation-corrected 
$\gamma$-ray energy of this event is $E_{\gamma} \simeq 3\times10^{48}$~erg, and 
it makes this event consistent with the Ghirlanda correlation 
\citep{ghirlanda04,ghisellini06a}.

For the prompt emission, we have analyzed the BAT and XRT data and 
found that the non-thermal component of the prompt emission may be fitted by the 
Band function similarly to other GRBs.
The low- and high-energy photon indices are almost constant with 
$\alpha_B \simeq -1$ and $\beta_B \simeq -2.5$, respectively, which are quite
typical features of GRBs.
The $15-150$~keV flux of the prompt non-thermal emission shows a relatively shallow
decay, and we show that this decay
is not due to the curvature effect of the emitting shell that
ceases the emission process suddenly.
The decay of the non-thermal emission may directly
connect to the anomalous X-ray afterglow
detected up to $t \simeq 10^6$~s.
If this is correct, the central engine might be active for $\gtrsim 10^6$~s.
In summary, the low-luminosity GRB 060218 has a typical prompt non-thermal
emission and may originate from a standard collapsar jet possibly driven by 
a long-acting central engine.

A number of {\it Swift} GRBs show late-time X-ray flares, which may be attributed
to the long intermittent activities of their central engines
\citep[e.g.,][]{zhang06,ioka05}.
In contrast, the engine of this burst might has the long power-law activity.
Then it is possible that the engine of this burst is different from those
of typical GRBs which are believed to be black holes.
\citet{mazzali06} have performed a detailed modeling of the spectra and light
curve of the SN component and argued that the progenitor star of this event
had a smaller mass than other GRB-SNe, suggesting that a neutron star
rather than a black hole was formed as the central engine of the jet.
The intrinsic rate of such low-luminosity GRBs would be larger than the 
local rate of typical cosmological GRBs \citep{soderberg06b,pian06,cobb06,liang06b}.
For these reasons, low-luminosity GRBs might be a distinct
GRB population involving neutron star engines.
We speculate that massive progenitor stars form black holes at the 
core collapse which produce highly-relativistic jets making
high-luminosity GRBs with strong spiky prompt emissions and flares, 
while less massive progenitor stars form neutron stars which produce 
mildly-relativistic jets making low-luminosity GRBs with weak smooth prompt emissions.

If the opening angles of the low-luminosity GRBs are around $\sim 0.3$,
the true rate of the low-luminosity GRBs with a beaming correction is
$R_{\rm LL} \sim 10^3~{\rm Gpc}^{-3}~{\rm yr}^{-1}$. 
The local rate of Type Ibc SNe is $R_{\rm SN} \sim 10^4~{\rm Gpc}^{-3}~{\rm yr}^{-1}$ 
\citep{soderberg06b,dahlen04,cappellaro99}.
Then the low-luminosity GRBs might be created at $\sim 10 \%$ rate of Type Ibc SNe.
By comparison, the collimation-corrected rate of the typical cosmological GRBs is 
$\sim 1 \%$ of that of Type Ibc SNe. 
Note that it is also suggested that the birthrate of Galactic magnetars
is $\sim 10\%$ of SN rate \citep{kouve98}.

\begin{figure}
\plotone{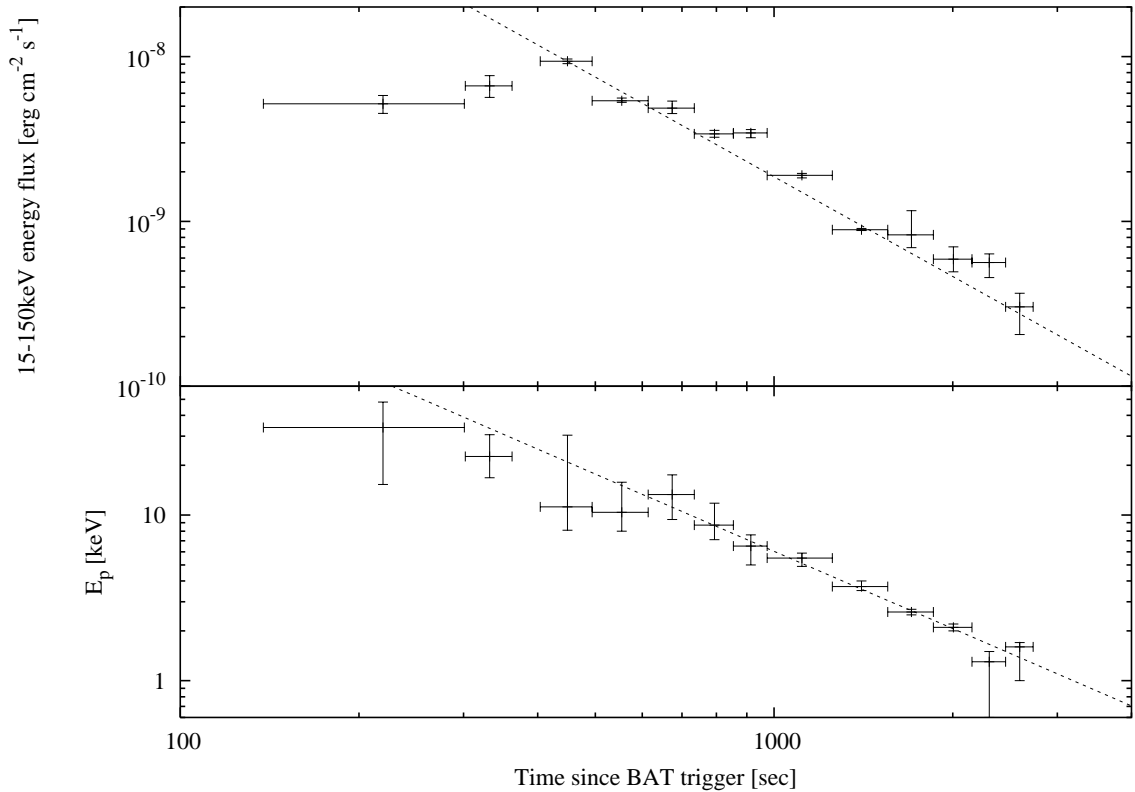}
\caption{
The light curve in 15-150~keV band and the temporal variation of the
spectral peak energy of the prompt non-thermal emission of GRB 060218,
from the joint fit of the BAT and XRT data by the absorbed Band function plus
the blackbody function.
The dotted lines show the best-fit power-law decays excluding the first two time bins
$F(t) \propto t^{-2.0}$ and $E_p(t) \propto t^{-1.6}$.
}
\label{fig:fluxep_l}
\end{figure}

\begin{figure}
\plotone{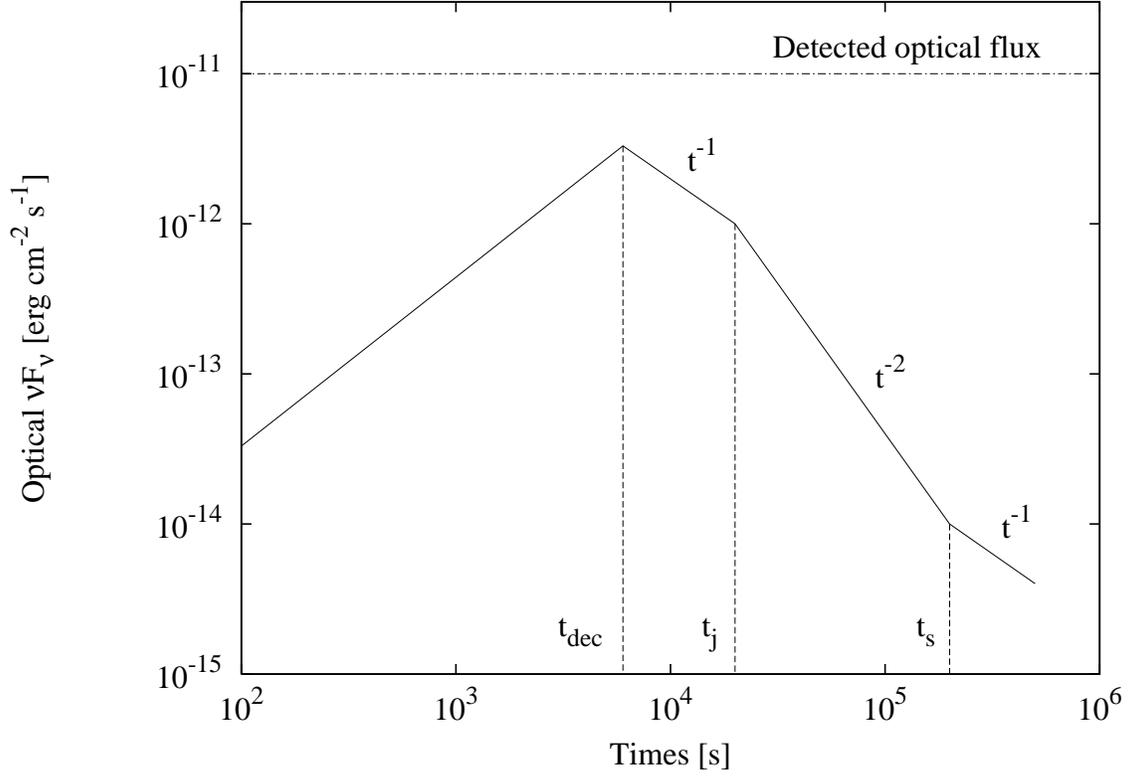}
\caption{
Schematic picture of the optical $\nu F_{\nu}$ flux from the external shock
in our jet model ({\it solid line}), which does not exceed the detected optical
flux shown in \citet{ghisellini06b} and \citet{campana06} ({\it dot-dashed line}).
The outflow in our jet model is decelerated at $t_{\dec}$ followed by the 
Blandford-McKee evolution $F_{\rm opt} \propto t^{-1}$, begins to expand sideways
at $t_j$ followed by the evolution $F_{\rm opt} \propto t^{-2}$, and finally shifts 
at $t_s$ into the Sedov-Tailor evolution $F_{\rm opt} \propto t^{-1}$.
}
\label{fig:optmodel}
\end{figure}

Now if the neutron star loses its rotational energy mainly through 
magnetic dipole radiation, and a fraction of energy $f$ is transferred to the jet, 
then the luminosity of the jet is approximated by \citep{shapiro83}
\begin{equation}
L_j \sim f L_0 \left(1+\frac{t}{\tau}\right)^{-2},
\end{equation}
where
\begin{eqnarray}
L_0 = \frac{B^2 {R_s}^6 {\Omega_0}^4}{6 c^3} 
\sim 10^{47}~{\rm erg}~{\rm s}^{-1}~{B_{16}}^2 {R_{s,6}}^6 {P_{0,-2}}^{-4}, \\
\tau = \frac{3 c^3 I}{B^2 {R_s}^6 {\Omega_0}^2}
\sim 10^3~{\rm s}~I_{45} {B_{16}}^{-2} {R_{s,6}}^{-6} {P_{0,-2}}^2.
\end{eqnarray}
Here $B$ is the magnetic field strength at the pole, $R_s$ is the neutron star
radius, $\Omega_0$ is the initial angular frequency, $P_0 = 2\pi/\Omega_0$ is the 
initial rotation period, and $I$ is the moment of inertia of the neutron star.
The characteristic spin-down timescale $\tau$ might be the peak time $t \sim 10^3$~s
of the prompt non-thermal emission, and the temporal decay of the spin-down
luminosity after $\tau$, $L_j \propto t^{-2}$, might agree with that of
the prompt non-thermal emission flux, $F \propto t^{-2.0}$, 
shown in Figure~\ref{fig:fluxep_l}.
Assuming $f \sim 10^{-2}$, we obtain $B \sim 10^{16}$~G and $P_0 \sim 10$~ms 
in order that $L_j \sim 10^{45}~{\rm erg}~{\rm s}^{-1}$ and $\tau \sim 10^3$~s are 
reproduced.
In this case, the rotational energy $I \Omega^2/2$ would be overwhelmed by 
the magnetic energy $(B^2/8\pi)(4\pi{R_s}^3/3)$ at $t \sim 10^4$~s.
This timescale is comparable to the transition time of the decay index of the prompt 
non-thermal emission from $F \propto t^{-2.0}$ to $F_{X} \propto t^{-1.1}$.
If this scenario was true, the rotation period would evolve as 
$P \sim 10(t/\tau)^{1/2}$~ms and become $P \sim 1$~s at $t \sim 1$~yr.
If a giant flare occurs at this magnetar like
soft gamma-ray repeater (SGR) 1806-20 \citep{hurley05},
this would be a clear evidence for our proposal.

The emission mechanism of the prompt non-thermal emission remains unclear.
\citet{ghisellini06b} have drawn the de-absorbed $\nu F_{\nu}$ spectrum from the 
{\it Swift} UltraViolet/Optical Telescope (UVOT) data and argued that the extrapolation
of the non-thermal spectrum with a constant low-energy photon index to the optical band 
joins with the detected thermal optical flux.
This implies that the non-thermal spectrum does not extend below 
$\sim 10^{15}$~Hz.
If this cutoff is due to the synchrotron self-absorption in the emitting shell with
the isotropic kinetic energy $L_{\iso} \sim 10^{47}~{\rm erg}~{\rm s}^{-1}$ 
and the bulk Lorentz factor $\Gamma_0 \simeq 5$, the emission radius of
the non-thermal emission is estimated as $r_0 \sim 10^{12}$~cm.
It is less than the deceleration radius of the afterglow shell estimated by
$r_{\dec} \simeq 4{\Gamma_0}^2 ct_{\dec} \simeq 2\times10^{16}$~cm in our jet model.
This implies that the prompt non-thermal emission may be produced by the internal 
dissipation of the jet.
\citet{ghisellini06b} have shown that a synchrotron inverse-Compton model could
reproduce the prompt non-thermal emission, the anomalous X-ray afterglow, and
the UV/optical thermal component with using parameters similar to those we suggest
in this paper, $\Gamma_0 = 5$, $\theta_0 = 0.2$, and $r_0 = 7\times10^{11}$~cm.

The origins of the thermal components of this event are also unclear.
A possible candidate is the emission from the expanding cocoon
as suggested by \citet{fan06} and \citet{ghisellini06b}
\citep[see also][]{ramirez02}.
Within the hot jet model discussed in \S~\ref{sec:collapsar}, the energy deposited into
the cocoon by the breakout time is estimated by $L_j t_{\br} \sim 3\times10^{47}$~erg,
which is smaller than the energy of the detected thermal emission $\gtrsim 10^{49}$~erg.
The rest  energy in the cocoon is estimated by 
$M c^2 {\theta_{\br}}^2/4 \sim 10^{49}$~erg, so that the cocoon expansion velocity
becomes $\beta_c \sim 10^{-1}$ outside the star.
The cocoon shell catches up with the afterglow shell at 
$t \sim r_{\dec}/c\beta_c \sim 10^7$~s, and thus the cocoon energy will not
affect the radio afterglow.
The thermal components are detected in the X-ray band at $t \lesssim 10^4$~s and in
the UV/optical band at $10^4 \lesssim t \lesssim 10^5$~s.
To explain such a behavior is an interesting future problem.

\acknowledgements
We appreciate the comments from the referee. 
We thank A.~Mizuta, K.~Murase, and N.~Kawanaka for useful discussions.
This work is supported in part by
Grant-in-Aid for the 21st Century COE
``Center for Diversity and Universality in Physics''
from the Ministry of Education, Culture, Sports, Science and Technology
(MEXT) of Japan.
K.~T. was supported by the JSPS Research Fellowship for Young Scientists,
grant 182666.
K.~I. was supported by the Grant-in-Aid (18740147) from the MEXT of Japan.
T.~S. was supported by an appointment of the NASA Postdoctoral Program
at the Goddard Space Flight Center, administered by Oak Ridge Associated
Universities through a contract with NASA.

\clearpage
%%%%%%%%%%%%%%
\begin{deluxetable}{lccccccc}
\tablewidth{0pt}
\tablecaption{
Spectral parameters of the non-thermal component of GRB 060218 (1)
\label{tab:band}
}
\tabletypesize{\scriptsize}
\tablehead{
  \colhead{$t$ (s)} &
  \colhead{time width (s)} &
  \colhead{$\alpha_B$} &
  \colhead{$\beta_B$} &
  \colhead{$E_p$ (keV)} &
  \colhead{$\chi^2$} &
  \colhead{dof} &
  \colhead{$F_{15-150{\rm keV}}$ ($10^{12}{\rm erg}{\rm cm}^{-2}$${\rm s}^{-1}$)} \\
}
\startdata
219.5 & 81.5 & $-1.25_{-0.05}^{+0.05}$ & $-3.8_{-6.2}^{+1.9}$ & $34_{-19}^{+14}$ 
& 232.7 & 292 & $5170_{-640}^{+640}$ \\
332.0 & 30.0 & $-1.47_{-0.07}^{+0.08}$ & $-8.4$ & $23_{-6}^{+8}$
& 156.7 & 190 & $6650_{-990}^{+1020}$ \\
449.0 & 45.0 & $-1.08_{-0.06}^{+0.06}$ & $-2.1_{-0.2}^{+0.2}$ & $11_{-3}^{+19}$
& 239.4 & 297 & $9370_{-320}^{+310}$ \\
554.0 & 60.0 & $-1.05_{-0.05}^{+0.05}$ & $-2.4_{-0.3}^{+0.2}$ & $10_{-2}^{+5}$
& 287.8 & 385 & $5400_{-120}^{+210}$ \\
674.0 & 60.0 & $-1.07_{-0.06}^{+0.07}$ & $-2.8_{-0.6}^{+0.3}$ & $13_{-4}^{+4}$
& 391.4 & 403 & $4870_{-360}^{+510}$ \\
794.0 & 60.0 & $-1.04_{-0.05}^{+0.05}$ & $-2.6_{-0.4}^{+0.3}$ & $9_{-2}^{+3}$
& 390.7 & 409 & $3400_{-150}^{+170}$ \\
914.0 & 60.0 & $-1.07_{-0.04}^{+0.06}$ & $-2.5_{-0.4}^{+0.3}$ & $6.5_{-1.5}^{+1.1}$
& 388.5 & 429 & $3450_{-230}^{+160}$ \\
1114.0 & 140.0 & $-1.17_{-0.03}^{+0.06}$ & $-3.0_{-0.9}^{+0.2}$ & $5.5_{-0.6}^{+0.4}$
& 629.0 & 573 & $1903_{-66}^{+52}$ \\
1404.0 & 150.0 & $-1.31_{-0.04}^{+0.10}$ & $-4.0_{-5.2}^{+0.8}$ & $3.7_{-0.2}^{+0.3}$
& 576.8 & 546 & $890_{-10}^{+17}$ \\
1704.5 & 150.5 & $-1.1_{-0.2}^{+0.2}$ & $-2.6_{-0.2}^{+0.2}$ & $2.6_{-0.1}^{+0.1}$
& 438.9 & 486 & $830_{-140}^{+330}$ \\
2005.0 & 150.0 & $-1.1_{-0.2}^{+0.5}$ & $-2.6_{-0.1}^{+0.1}$ & $2.1_{-0.1}^{+0.1}$
& 408.6 & 437 & $590_{-100}^{+110}$ \\
2305.0 & 150.0 & $-1.1_{-0.5}^{+0.5}$ & $-2.53_{-0.05}^{+0.05}$ & $1.3_{-1.2}^{+0.2}$
& 375.8 & 393 & $563_{-106}^{+72}$ \\
2593.5 & 138.5 & $-1.0_{-0.4}^{+0.5}$ & $-2.8_{-0.1}^{+0.1}$ & $1.6_{-0.6}^{+0.1}$
& 300.9 & 344 & $303_{-98}^{+63}$ \\
\enddata
\tablecomments{
These values are calculated from the joint fit of the BAT and XRT data
by the absorbed Band function plus the blackbody function.
Errors are for 90\% confidence.
}
\end{deluxetable}
%%%%%%%%%%%%%%
\begin{deluxetable}{lccccc}
\tablewidth{0pt}
\tablecaption{
Spectral parameters of the non-thermal component of GRB 060218 (2)
\label{tab:cpl}
}
\tabletypesize{\scriptsize}
\tablehead{
  \colhead{$t$ (s)} &
  \colhead{time width (s)} &
  \colhead{$\alpha_c$} &
  \colhead{$E_p$ (keV)} &
  \colhead{$\chi^2$} &
  \colhead{dof} \\
}
\startdata
219.5 & 81.5 & $-1.36_{-0.05}^{+0.04}$ & $36_{-8}^{+15}$ & 230.0 & 293 \\
332.0 & 30.0 & $-1.46_{-0.08}^{+0.08}$ & $22_{-5}^{+8}$ & 156.6 & 191 \\
449.0 & 45.0 & $-1.39_{-0.05}^{+0.05}$ & $31_{-7}^{+11}$ & 237.8 & 298 \\
554.0 & 60.0 & $-1.34_{-0.05}^{+0.05}$ & $22_{-5}^{+7}$ & 293.0 & 386 \\
674.0 & 60.0 & $-1.18_{-0.05}^{+0.06}$ & $16_{-3}^{+4}$ & 381.2 & 404 \\
794.0 & 60.0 & $-1.40_{-0.06}^{+0.06}$ & $15_{-4}^{+7}$ & 393.2 & 410 \\
914.0 & 60.0 & $-1.44_{-0.06}^{+0.07}$ & $12_{-3}^{+5}$ & 400.2 & 430 \\
1114.0 & 140.0 & $-1.42_{-0.07}^{+0.07}$ & $5.7_{-0.6}^{+0.8}$ & 626.8 & 574 \\
1404.0 & 150.0 & $-1.38_{-0.08}^{+0.09}$ & $3.6_{-0.2}^{+0.1}$ & 573.9 & 547 \\
1704.5 & 150.5 & $-1.6_{-0.1}^{+0.1}$ & $2.2_{-0.1}^{+0.1}$ & 448.5 & 487 \\
2005.0 & 150.0 & $-1.8_{-0.2}^{+0.1}$ & $1.1_{-0.1}^{+0.2}$ & 420.3 & 438 \\
2305.0 & 150.0 & $-1.9_{-0.1}^{+0.1}$ & $0.7_{-0.1}^{+0.1}$ & 392.8 & 394 \\
2593.5 & 138.5 & $-1.7$ & $0.5$ & 314.7 & 345 \\
\enddata
\tablecomments{
These values are calculated from the joint fit of the BAT and XRT data
by the absorbed CPL model plus the blackbody function.
Errors are for 90\% confidence.
}
\end{deluxetable}
%%%%%%%%%%%%%%
%%%%%%%%%%%%%%
%%%%%%%%%%%%
%%%%%%%%%%%%%%
%%%%%%%%%%%%
%%%%%%%%%%%%%%
%%%%%%%%%%%%

\end{document}